\documentstyle[12pt,world_sci]{article}
\include{psfig}   
\begin{document}
\title{{\bf SEARCH FOR FIRST AND SECOND GENERATION
LEPTOQUARKS  AT D\O \    \thanks{Submitted to the proceedings of the 1994
meeting of the Division of Particles and Fields in
Albuquerque NM.}}} \author{DOUGLAS  NORMAN\thanks{Presented for the D\O \
Collaboration at DPF '94} \\ {\em Texas A\&M University,
College Station, TX  77843}}
\maketitle
\setlength{\baselineskip}{2.6ex}
\begin{center}
\parbox{13.0cm}
{\begin{center} ABSTRACT \end{center}
{\small \hspace*{0.3cm} A search for first and  second generation leptoquarks
has been done with the D\O \ detector at
Fermilab's  p\={p} collider with $\sqrt{s}=1.8$ TeV.  95\% C.L. mass limits for
first generation scalar leptoquarks have been
recently published. The number for the total integrated luminosity used in the
first generation leptoquark analysis has changed
(via a change in the total inelastic cross section) since the publication. The
new limits are 130 GeV/c$^{2}$ and 116 GeV/c$^{2}$
for a respective 100\% and 50\% decay branching ratio of the leptoquark to
electron. The preliminary upper limit on the cross
section from the search for second generation scalar leptoquarks has set limits
on the mass of the second generation leptoquark
of 97 GeV/c$^{2}$ for 100\% branching to muons and 80 GeV/c$^{2}$ for 50\%
branching. In contrast with leptoquark detection
thresholds at e$^{+}$e$^{-}$ and e-p machines, these limits are independent of
the unknown coupling of the leptoquark to leptons
and quarks.}}
\end{center}

\section{Introduction} Leptoquarks are exotic particles with both color and
lepton quantum numbers. They
are bosons that appear as spin = 0 or spin = 1 particles in many SUSY, GUT and
composite models~\cite{theory}. The coupling
constant  of leptoquarks to leptons and quarks is an unknown parameter,
$\lambda$, in  the  theory,  but  experimental
constraints~\cite{eeconst} require the coupling, $\lambda ^{2}/4\pi$, to be on
the order 0.1$\alpha  _{em}$ for masses of the
leptoquark that can be probed with the current D\O \ data.  Since leptoquarks
have color, they can be produced via the strong
interaction as leptoquark - anti-leptoquark pairs.  For light leptoquarks (
$\sim$\llap{$^<$} 1-100 TeV) the coupling of the
leptoquark to fermions is required to be generational; for example, a first
generation leptoquark may only couple to electrons,
electron neutrinos, u and d quarks. This restriction is required to prevent
leptoquarks from contributing to the violation of the
limits on the branching ratio of rare decays such as $K^{+}  \rightarrow
e^{+}\nu$.  The expected signatures for light first
generation leptoquark pairs are: two electrons plus at least two jets, one
electron plus missing Et plus at least two jets, and
missing Et plus two or more jets. The expected signatures for a second
generation leptoquark pair are the same except the
electrons are replaced by muons. The leptons, missing Et, and jets in
leptoquark events are also expected to be well isolated from
each other in general.

\section{Analysis} A description of the D\O \ detector is given
elsewhere~\cite{det}. Also, details of the first generation analysis can be
found in a recent publication~\cite{firstlq}. This report will cover the
analysis for the second generation leptoquark search. Only those signatures
which contain at least one muon are dealt with here.

The data used in this analysis were taken between September of 1992 and
May of 1993 during the 1992-1993 Tevatron collider run. It represents
12.3 pb$^{-1}$.
Two data sets were selected, one for the di-muon leptoquark signature and
one for the single muon leptoquark signature. Both data sets were required to
pass a
trigger with a muon and jet requirement.
The hardware portion of the trigger required a muon with Et greater than 3 GeV
and $\mid \eta \mid < 2.4$, and one jet tower ($0.2\times 0.2$ radians in $\eta
\times
\phi$) with Et greater than 5 GeV. In the software portion of the trigger,
one muon with Et greater than 8 GeV and one jet with Et greater than 15 GeV
were required. The trigger cut left 751 events in the two muon sample
and 2938 events in the single muon sample.

The event selection for the two muon leptoquark signature is given
in table 1. The event selection for the single muon leptoquark
signature
is given in table~\ref{tab:mu}. The muon quality cuts mentioned in both
tables 1 and ~\ref{tab:mu}
are for rejecting cosmic muons and also combinatorics which are reconstructed
out of stray hits in the muon chambers. The muon isolation cut mentioned in
table 1 requires that there is no jet within 0.65 radians of
the muon(s) in question. For the single muon selection no jet within 0.7
radians of the muon in question is allowed. The $\Delta \phi _{\mu \mu}$ cut
in table 1 is for eliminating low mass di-muon events such as
J/$\Psi$'s. Here $\phi$ is the azimuthal angle. The last event after the
jet Et cut in the two muon event sample is rejected by excluding events with
back to back muons in $\phi$. This cut rejects 75\%\ of the Z background.
 The expected number of background for the cuts used in the two muon
event selection was estimated to be about 0.9 events
from Drell-Yan, b\={b} production, and fake muons from punchthrough of pions in
QCD events.
\begin{table}[h]
\label{tab:mu2}
\vspace{-0.2cm}
\caption{The event selection for the two muon leptoquark signature}
\hspace{1.2cm}
\begin{tabular}{|l|c|}
\hline
\hline
Selection cut & Number of events surviving cut \\
\hline
Trigger selection & 751 \\
\hline
Two muons with Pt $>$ 25 GeV/c & 206 \\
\hline
Two muons: $\mid \eta \mid < 1.7$; one muon: $\mid \eta \mid < 1.0$ & 102 \\
\hline
One muon passes muon quality cuts & 49 \\
\hline
Two muons pass muon isolation cut & 18  \\
\hline
$\Delta \phi _{\mu \mu} < 0.1$ rad & 17 \\
\hline
Two jets Et $>$ 25 GeV & 1 \\
\hline
$\mid \pi - \Delta  \phi _{\mu \mu}\mid > 0.2$ rad & 0 \\
\hline
\end{tabular}
\end{table}

The detector clean up cuts in the single muon analysis are used
to eliminate those events that have badly measured jets or large amounts
of electronic noise in the calorimeter which are more problematical in
measuring missing Et.
The back to back cut between the muon and the missing Et,
$\mid \pi - \Delta  \phi _{\mu ,MEt}\mid > 0.2$ rad, in table~\ref{tab:mu} is
designed to eliminate W associated events and also some
events with badly measured muon momentum, since these events are expected
to have the missing Et back to back in $\phi$ with the muon.
The number of expected background events for the cuts given in
table~\ref{tab:mu} was estimated
to be about 1.1 events from
$W\rightarrow \mu \nu$
plus jets, b\={b} production, and fake muons from punchthrough.
\vspace{-0.4cm}
\begin{table}[h]
\label{tab:mu}
\caption{The event selection for the single muon leptoquark signature}
\vspace{0.1cm}
\hspace{2cm}
\begin{tabular}{|l|c|}
\hline
\hline
Trigger selection & 2938 \\
\hline
Detector cleanup & 1925 \\
\hline
One muon Pt $>$ 20 GeV/c & 1660 \\
\hline
One muon: $\mid \eta \mid < 1.0$ & 1232 \\
\hline
One muon passes muon quality cuts & 400 \\
\hline
Missing Et $>$ 25 GeV leaves  & 295 \\
\hline
One muon passes isolation cut & 92 \\
\hline
$\mid \pi - \Delta \phi _{\mu, MEt}\mid > 0.2$ rad & 59 \\
\hline
Two jets Et $>$ 25 GeV & 18 \\
\hline
Transverse mass for muon and missing Et $>$ 95 GeV/c$^{2}$ & 0 \\
\hline
\end{tabular}
\end{table}

 With no observed events, the upper limit on the cross section was calculated
as in equation~\ref{eqn:cross}.
\begin{eqnarray}
\beta ^{2} \times \sigma ^{\mu \mu} & = & N_{\mu \mu}
^{95\%\ CL}
/\left(\epsilon _{\mu \mu} \cdot L\right) \nonumber \\
\vspace{.2in}
2\beta \left(1-\beta \right) \times \sigma ^{\mu \nu} & = &
N_{\mu \nu}^{95\%\ CL}
/\left(\epsilon _{\mu \nu} \cdot L\right) \label{eqn:cross}
\end{eqnarray}
Here $\beta$ is the branching fraction for leptoquark decay to muon plus quark.
The total efficiency for the two signatures is represented by
$\epsilon _{\mu \mu}$ and $\epsilon _{\mu \nu}$. L is the integrated luminosity
which was 12.3 pb$^{-1}$ for this analysis.
The kinematic efficiencies for detecting leptoquarks were determined from
ISAJET~\cite{isajet} Monte Carlo processed with GEANT~\cite{geant}. The
muon detection efficiency was determined from a study of the data.
And the trigger efficiency was determined from the data and a study
of the leptoquark Monte Carlo processed with a simulation of the D\O \ trigger.
The total preliminary efficiency for the two muon plus two jet leptoquark
signature
varied from 0.27\%\ to 7.1\%\ for masses ranging from 45 to
250 GeV/c$^{2}$. For the single muon plus missing Et plus two
jet signature,
the total efficiency varied from 0.1\%\ to 4.56\%\ for masses
ranging from 45 to 200 GeV/c$^{2}$.

The number of expected events with no events seen was calculated according to
the method by Cousins and Highland~\cite{cousins}
using the total uncertainty on $\epsilon \cdot L$ given in
table~\ref{tab:error}.
These uncertainties are the total systematic and statistical uncertainties
added in
quadrature. The preliminary 95\%\ CL limits on the number of expected events,
$N^{95\%\ CL}_{\mu \mu}$ and $N^{95\%\ CL}_{\mu \nu}$,
are also given for each mass in table~\ref{tab:error} along with the upper
limit on the measured cross section times the appropriate branching ratio
factors.
\begin{table}
\caption{ The preliminary total uncertainty (Err) from the second generation
analysis
and the 95\%\ CL for the upper limit on the number of expect events and
measured
cross section}
\label{tab:error}
\hspace{2.5cm}
\begin{tabular}{|c|c|c|c|c|c|c|}
\hline
\hline
Mass (GeV/c$^{2}$) & 45 & 75 & 100 & 150
& 200  & 250  \\
\hline
\hline
Err$_{\mu \mu}(\%)$ & 39.8 & 30.8 & 29.9 & 28.7 & 29.0 & 28.9 \\
\hline
Err$_{\mu \nu}(\%)$ & 24.3 & 19.7 & 19.3 & 19.9 & 20.0 & - \\
\hline
$N^{95\%\ CL}_{\mu \mu}$ & 4.87 & 3.62 & 3.57 & 3.51 & 3.52 & 3.51 \\
\hline
$N^{95\%\ CL}_{\mu \nu}$ & 3.32 & 3.19 & 3.18 & 3.2 & 3.22 & - \\
\hline
$\beta ^{2} \times \sigma ^{\mu \mu}$(pb) &149 & 14.8 & 8.34 & 4.62 & 4.28
& 4.02 \\
\hline
$2\beta(1-\beta)\times \sigma ^{\mu \nu}$(pb) & 261 & 32.6 & 14.3 & 8.5 & 5.74
&
- \\
\hline
\end{tabular}
\end{table}
\section{Results and Conclusion}
\begin{figure}[h]
 \vspace{-0.7cm}
   \centerline{\psfig{figure=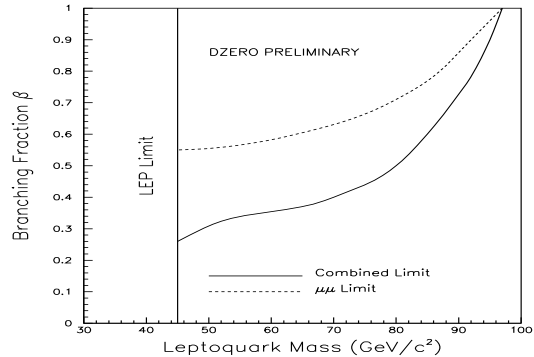,height=2.1in,width=2.4in}}
 \vspace{-0.2cm}
  \caption{The D\O \ preliminary 95\%\ CL excluded region of branching
fraction vs leptoquark mass}
  \label{fig:bvm}
\end{figure}

Comparing the measured upper limit on the cross section with the
theoretical cross section from ISAJET with Morfin and Tung leading order
parton distribution functions, the D\O \
preliminary mass limit
from the combined signal for 100\%\ branching of leptoquark to muons was
determined to be 97 GeV/c$^{2}$. For 50\%\ branching, the mass limit was
80 GeV/c$^{2}$. The branching fraction vs. leptoquark mass excluded region
is given in
figure~\ref{fig:bvm}.

The D\O \ total integrated luminosity has changed since the publication of
reference 4 due to an update of the total inelastic cross section
(averaged new CDF and E710)
used to
calculate the luminosity.
The new luminosity is 13.4 pb$^{-1}$ for the
first generation leptoquark search.
The mass limits for first generation leptoquarks are now 130 GeV/c$^{2}$ for
100\%\
branching of leptoquark to electron and 116 GeV/c$^{2}$ for 50\%\ branching.

\end{document}